\documentclass[lettersize,journal]{IEEEtran}
\usepackage{amsmath,amsfonts}
\usepackage{algorithmic}
\usepackage{algorithm}
\usepackage{array}
\usepackage[caption=false,font=scriptsize,labelfont=sf,textfont=sf]{subfig}
\usepackage{textcomp}
\usepackage{stfloats}
\usepackage{url}
\usepackage{verbatim}
\usepackage{graphicx}
\usepackage{cite}
\usepackage{amssymb}
\usepackage{bm}
\usepackage{cuted}
\usepackage{placeins}

\usepackage[utf8]{inputenc}
\usepackage{lineno} 
\usepackage{cases}
\usepackage{float}  
\usepackage{textcomp}
\usepackage{bbold}
\usepackage{xcolor}
\usepackage{siunitx}

\usepackage[caption=false,font=scriptsize]{subfig}
\usepackage{amsthm}
\newtheorem{theorem}{Theorem}
\newtheorem{corollary}{Corollary}

\begin{document}
	
	\bstctlcite{BSTcontrol}

\title{Sum Rate Maximization for Practical RIS-Enhanced Rate-Splitting Multiple Access Systems}

\author{Xuejun Cheng,~\IEEEmembership{Graduate Student Member,~IEEE}, Qian Zhang,~\IEEEmembership{Member,~IEEE}, Jiaming Shi, Yufei Zhao, \\ Rugui Yao,~\IEEEmembership{Senior Member,~IEEE}, Ye Fan,~\IEEEmembership{Member,~IEEE}
\thanks{\scriptsize This research was supported in part by the Key Research and Development Project in Shaanxi Province under Grant 2025CY-YBXM-055 and Grant 2025CG-GJHX-13.}

\thanks{\scriptsize Xuejun Cheng and Qian Zhang are with School of Computer and Communication Engineering, Northeastern University at Qinhuangdao, Qinhuangdao 066004, China (e-mail: chegnxuejun@mail.sdu.edu.cn, zhangqian@neuq.edu.cn). Xuejun Cheng is also with School of Information Science and Engineering, Shandong University, Qingdao 266237, China.}
\thanks{\scriptsize Jiaming Shi is with the Department of Electrical and Computer Engineering, National University of Singapore, Singapore 117575 (e-mail: shij@u.nus.edu).}
\thanks{\scriptsize Yufei Zhao is with the School of Electrical and Electronics Engineering, Nanyang Technological University, Singapore 639798 (e-mail: yufei.zhao@ntu.edu.sg).}
\thanks{\scriptsize Rugui Yao and Ye Fan are with the School of Electronics and Information, Northwestern Polytechnical University, Xi’an, Shaanxi 710072 China (e-mail: yaorg@nwpu.edu.cn, fanye@nwpu.edu.cn). Ye Fan is also with the Shenzhen Research Institute of Northwestern Polytechnical University, Shenzhen 518057, China.}}

\maketitle

\begin{abstract}
Reconfigurable intelligent surfaces (RIS) combined with rate-splitting multiple access (RSMA) provide a promising solution for improving the spectral efficiency (SE) of sixth-generation (6G) communication systems. However, most existing studies neglect the practical coupling between the RIS phase shift and amplitude, which may cause phase-shift mismatch and significant performance loss. 
In this paper, we investigate the multiuser sum-rate maximization problem for practical RIS-enhanced RSMA systems considering amplitude-phase coupling. 
To address the resulting non-convex problem, an effective variable separation algorithm is proposed. 
The proposed algorithm supports arbitrary amplitude-phase coupling relationships and both continuous and discrete RIS phase shifts, and can also be directly extended to practical RIS-enhanced NOMA and OMA systems.
Simulation results demonstrate sum-rate gains of over 190\% and 50\% over the ideal-RIS-based design applied to practical system and the continuous-to-1-bit quantization scheme, respectively, while the proposed RSMA scheme significantly outperforms NOMA and OMA.

\end{abstract}

\vspace{-0.1cm}
\begin{IEEEkeywords}
Practical reconfigurable intelligent surfaces; rate-splitting multiple access; practical amplitude response.
\end{IEEEkeywords}

\vspace{-0.5cm}
\section{Introduction}
\vspace{-6pt}
Sixth-generation (6G) wireless networks are expected to provide broader and more extensive coverage at low cost and with low energy consumption~\cite{Chen2026Standardization6G}. 
Reconfigurable intelligent surfaces (RIS) consist of a large number of low-cost passive reflecting elements and can reconfigure the wireless propagation environment by adjusting the phase shifts of reflected signals, thereby improving communication quality and coverage~\cite{zhang2025bdrisnoma,9,zhang2024activerissumrate}. Rate-splitting multiple access (RSMA) is regarded as an efficient and reliable multiple-access and interference management technique for 6G wireless networks~\cite{22}. By partially decoding interference while treating the remaining interference as noise, RSMA enables more flexible multiuser interference management. Existing studies have shown that RSMA can achieve higher spectral efficiency~\cite{mao2022rate} and energy efficiency~\cite{yang2021optimization} than conventional schemes such as orthogonal multiple access (OMA) and non-orthogonal multiple access (NOMA).

In recent years, the integration of RIS and RSMA has attracted extensive attention~\cite{Salem2025}.
In~\cite{Chen2025RISRSMA}, Chen {\it et al.} jointly optimized beamforming and common-rate allocation for energy-efficient RIS-RSMA ISAC systems. 
In~\cite{Nduwayezu2026RISRSMA}, Nduwayezu {\it et al.} investigated antenna and RIS partitioning for sum-rate maximization in RSMA-aided vehicular networks.
However, most existing studies on RIS are based on the ideal phase-shift model, where the reflecting amplitude is generally assumed to remain constant~\cite{zhang2026risnomahwi}. In~\cite{Abeywickrama2020NonIdealRIS}, Abeywickrama {\it et al.} revealed the inherent coupling relationship between the amplitude and phase shift of RIS elements and proposed a corresponding general analytical model. In~\cite{zhang2024practical}, Zhang {\it et al.} proposed an amplitude-feedback method to obtain the practical amplitude-phase coupling characteristics of RIS elements.

To the best of our knowledge, existing studies on RIS-enhanced RSMA systems rarely consider amplitude-phase coupling. However, since the phase shifts optimized under the ideal RIS model must ultimately be applied to practical systems, neglecting this coupling may lead to significant performance degradation.
To address this issue, we investigate the multiuser sum-rate maximization problem for practical RIS-enhanced RSMA systems by considering amplitude-phase coupling.
To solve the resulting highly non-convex optimization problem, we propose a flexible variable-separation-based algorithm that supports different amplitude-phase coupling relationships and both continuous and 1-bit discrete phase shifts, and can be extended to practical RIS-enhanced NOMA and OMA systems.
Simulation results validate the effectiveness of the proposed scheme and optimization algorithm. 
Compared with the ideal-RIS-based design deployed on practical RIS system, the proposed scheme achieves performance gain of over 190\%.
Moreover, the practical RIS-enhanced RSMA system significantly outperforms the corresponding NOMA and OMA systems in terms of sum-rate performance.

\begin{figure}[t]
	\centering
	\includegraphics[width=0.4\textwidth]{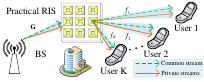}
	\vspace{-12pt}
	\caption{A Practical RIS-enhanced RSMA system.}
	\label{fig1}
	\vspace{-14pt}
\end{figure}

\vspace{-8pt}
\section{System Model and Problem Formulation}
\vspace{-2pt}
\label{s2}
We consider a downlink scenario in practical RIS-enhanced RSMA system, as shown in Fig.~\ref{fig1}. In this system, the BS is equipped with $M$ antennas using the linear precoding technique to serve $K$ single-antenna users with the assistance of a practical RIS equipped with $N$ reflecting elements, where the direct links between the BS and users are obstructed. 

\vspace{-15pt}
\subsection{Practical RIS Model}
\vspace{-4pt}

In the practical RIS system, there is a property for RIS, i.e., the phase-shift is coupled with the amplitude response.
We use the parallel resonant circuit-based model considered in \cite{Abeywickrama2020NonIdealRIS} to characterize the amplitude-phase model of the practical RIS.
Specifically, the amplitude response of RIS element $n$, denoted by $\beta_n$, varies with its phase-shift $\theta_n$ according to
\vspace{-7pt}
\begin{equation} \label{phase_amplitude}
		\beta_n(\theta_n) = \eta \left( {\rm sin}(\theta_n - \delta) + 1 \right)^{\alpha} + \beta_{\rm min},
		\vspace{-5pt}
\end{equation}
\vspace{-15pt}

\noindent  where $\eta = (1-\beta_{\rm min})/2^{\alpha}$. The values of the constants $\beta_{\rm min}\geq 0$, $\delta\geq 0$, and $\alpha\geq 0$ depend on the specific circuit parameters for the equivalent impedance of the RIS reflecting elements.
These parameters can be obtained from measured amplitude-phase responses using a standard curve-fitting tool~\cite{Abeywickrama2020NonIdealRIS}.

\vspace{-12pt}
\subsection{Signal Model}
\vspace{-3pt}

Based on the principle of rate splitting technique~\cite{22}, the signal transmitted by the BS is split into common signals and private signals. 
All common signals will be combined as a common stream $s_c$ and allowed to be decoded by every user. Meanwhile, the private signals will be encoded into private stream $\left\{s_i\right\}_{i=1}^K$, which will be allowed to be decoded only by their corresponding users. 
Therefore, the signal transmitted by the BS is
$\bm x \!=\! \bm w_cs_c \!+\! \sum_{i=1}^{K}{\bm w_i s_i},$
\noindent where $\bm w_c \!\in\!\mathbb{C}^{M \!\times\! 1}$ and $\bm \left\{\bm w_i\right\}_{i=1}^K \!\in\!\mathbb{C}^{M \!\times\! 1}$ denote the linear precoding vectors for the common stream and the private stream of user $k$, respectively. Define $\bm s \!=\! \left[s_c,s_1,s_2,\cdots,s_K\right]^\mathrm{T}$, where $s_c,s_k \!\sim\! \mathcal{C} \mathcal{N}\left(0,1\right)$ and $\mathbb{E} \left[\bm s \bm s^\mathrm{H}\right] = \mathbf{I}$, $k \in \mathcal{K} \triangleq \left\{1,2,\ldots, K\right\}$.

Then, the received signal at user $k$ can be represented as
\vspace{-8pt}
\begin{equation} 
	y_k = {\bm{h}_k^\mathrm{H} \bm{w}_c s_c}
	+ {\bm{h}_k^\mathrm{H} \bm{w}_k s_k}
	+ \sum\nolimits_{i=1,\, i \neq k}^{K}
	\bm{h}_k^\mathrm{H} \bm{w}_i s_i
	+ n_k,
\end{equation}
\vspace{-15pt}

\noindent where $\bm h_k^\mathrm{H} = \bm f_k^\mathrm{H} \bm\Theta \mathbf{G}$, $ n_k \sim \mathcal{C} \mathcal{N}\left(0,\sigma_k^2\right)$ denote the additive white Gaussian noise (AWGN) at user $k$.
$\mathbf{G}\in\mathbb{C}^{N\times M}$, $\bm f_k\in\mathbb{C}^{N\times 1}$ and $\bm \Theta = \mathrm{diag}\left(\bm \theta^\mathrm{H}\right)\in \mathbb{C}^{N \times N}$ denote the channel from BS to practical RIS, practical RIS to user $k$, and a phase-shift matrix of practical RIS, respectively.

Each user initially treats every private stream as interference and then decodes the common stream. After successful decoding, each user removes the common stream by successive interference cancellation (SIC) technique and then decodes its corresponding private stream. Hence, the signal-to-interference-plus-noise ratio (SINR) for the common and private stream of user-$k$ are given by
\vspace{-8pt}
\begin{equation*}
	\begin{split}
		\gamma_{c,k} \!=\! \frac{|\bm h_k^\mathrm{H} \bm w_c|^2}{\sum_{i=1}^{K}\!{|\bm h_k^\mathrm{H} \bm w_i|^2} \!+\!\sigma_k^2},
		\gamma_{k} \!=\! \frac{|\bm h_k^\mathrm{H} \bm w_k|^2}{\sum_{i=1,i\neq k}^{K}\!{|\bm h_k^\mathrm{H} \bm w_i|^2} \!+\!\sigma_k^2}.
	\end{split}
\end{equation*}
\vspace{-13pt}

The achievable rate of user-$k$ to decode the common stream is expressed as
$R_{c,k} = \log_2{\left(1+\gamma_{c,k}\right)}.$
To ensure that all users can successfully decode $s_c$, the common rate should not exceed $R_{c} = \min{\{R_{c,1},R_{c,2},\dots,R_{c,K}\}}$. 
Then, the achievable rate of user-$k$ to decode the desired private stream can be expressed as 
$R_k = \log_2{\left(1+\gamma_{k}\right)}.$ 
By combining the decoded common message with its corresponding private message, each user can reconstruct the original information. Then, the overall system sum rate is given by $R_{total} = R_c +\sum_{k=1}^{K} R_k$. 

\vspace{-12pt}
\subsection{Problem Formulation}
\vspace{-3pt}

In such a system, we aim to maximize the multiuser sum rate, where the practical RIS model is considered. The multiuser sum-rate maximization problem is formulated as
\vspace{-6pt}
\begin{subequations} 
	\label{problem_formulation} 
	\begin{align} 
		\max\limits_{\bm w,\bm \Theta} f&\!\left(\bm w,\bm \Theta\right)\! \!=\! \log_2\!\left(\!\!1 \!+\! \min\limits_{k \in \mathcal{K}}{\gamma_{c,k}}\!\right)\! \!+\!\! \sum\nolimits_{k=1}^{K}\! \log_2\!\left(\!1 \!+\! \gamma_{k}\!\right)\!,   \\ 
		\text{s.t.} & \mathcal{C}_{\text{BS}}: \|\bm{w}_c\|_2^2 + \sum\nolimits_{i=1}^{K}{\|\bm{w}_i\|_2^2} \leq P_{\max}, \\ 
		& \mathcal{C}_k : \gamma_{c,k} \geq \gamma_{th},\gamma_{k} \geq \gamma_{th}, \\ 
		& \mathcal{C}_{\theta}: -\pi \leq \theta_n \leq \pi,n=1,2,\cdots,N, \\ 
		& \mathcal{C}_{\beta}\!: \beta_n(\theta_n) = \eta \left( {\rm sin}(\theta_n - \delta) + 1 \right)^{\alpha} + \beta_{\rm min}, 
	\end{align} 
\end{subequations}
\vspace{-15pt}

\noindent where $\bm{w} = \left[ \bm{w}_c^\mathrm{T},\bm{w}_1^\mathrm{T},\bm{w}_2^\mathrm{T},\cdots,\bm{w}_K^\mathrm{T} \right]^\mathrm{T} \in \mathbb{C}^{M \left(K+1\right)\times1}$. 
$P_{\max}$ represents the maximum transmit power at the BS, and $\gamma_{th}$ represents the quality of service (QoS) threshold.


The optimization design of practical RIS systems differs significantly from that of ideal RIS systems, primarily due to the coupling between the reflection amplitude and phase shift. In the ideal RIS model, the reflection amplitude is typically assumed to be constant, whereas in practical RIS systems, it varies with the phase shift, resulting in a nonlinear amplitude-phase coupling constraint. Therefore, conventional unit-modulus-based optimization methods cannot be directly applied to practical RIS systems. Furthermore, in problem~\eqref{problem_formulation}, the coupling between the BS precoding vector $\bm w$ and the RIS reflection coefficient matrix $\mathbf{\Theta}$ further increases the non-convexity and difficulty of solving the optimization problem.

\vspace{-6pt}
\section{Beamforming Optimization}
\vspace{-2pt}
\label{s3}

In this section, we propose an effective algorithm to solve the problem \eqref{problem_formulation} by separating variables. To facilitate the subsequent optimization, auxiliary variables $\xi$ and $\{\xi_k\}_{k=1}^{K}$ are introduced, yielding an equivalent reformulation of problem~\eqref{problem_formulation} as follows:
\vspace{-6pt}
\begin{subequations}
	\label{problem_formulation1}
	\begin{align}
		\max\limits_{\bm w,\xi,\{\xi_k\}_{k=1}^K, \bm \Theta} \quad 
		& \mathcal{G} \!=\! \log_2\left(1 \!+\! {\xi}\right) 
		\!+\! \sum\nolimits_{k=1}^{K} \log_2\left(1 \!+\! \xi_k\right), \\
		\text{s.t.} \quad
		& \mathcal{C}_{\text{BS}},\mathcal{C}_{\theta},\mathcal{C}_{\beta}, 
		\mathcal{C}_{th} : \xi \geq \gamma_{th}, \xi_k \geq \gamma_{th}, \\
		& \mathcal{C}_k : \gamma_{c,k} \geq \xi, \gamma_{k} \geq \xi_k.
	\end{align}
\end{subequations}
\vspace{-18pt}

Owing to the coupling between $\bm w$ and $\bm\Theta$, we alternately update these two variable blocks while keeping the other one fixed. Accordingly, the resulting AO procedure is given by
\vspace{-5pt}
\begin{subequations}
	\label{AO}
	\begin{align}
		\label{AO-w}
		\bm w^{m+1} = \arg \max\limits_{\bm w,\xi,\xi_k} \mathcal{G}\quad &\text{s.t.} \mathcal{C}_{\text{BS}},\mathcal{C}_k,\mathcal{C}_{th},  \\
		\label{AO-theta}
		\mathbf{\Theta}^{m+1} = \arg \max\limits_{\bm \Theta , \xi,\xi_k} \mathcal{G}\quad &\text{s.t.} \mathcal{C}_k,\mathcal{C}_{th},\mathcal{C}_{\theta},\mathcal{C}_{\beta}. 
	\end{align}
\end{subequations}
\vspace{-18pt}

\subsection{Optimize Precoding Vector}
\vspace{-2pt}
With $\mathbf{\Theta}$ fixed, we can reformulate the objective function in the following form. 
\vspace{-8pt}
\begin{equation}
	\label{Precoding Vector}
		\max\limits_{\bm w,\xi,\{\xi_k\}_{k=1}^K} \ \mathcal{G}  \quad
		\text{s.t.} \ \mathcal{C}_{\text{BS}}, \mathcal{C}_{th}, \mathcal{C}_k. 
\end{equation}
\vspace{-5pt}

\begin{theorem}
	\label{theorem1}
	Given ratios $\frac{|A_k\left(\bm x\right)|^2}{B_k\left(\bm x\right)}$ with $A_k\left(\bm x\right) \in \mathcal{C}$ and $B_k\left(\bm x\right) > 0 $, the ratio problem is defined to be of the form:
	\vspace{-6pt}
	\begin{equation}
		\max_{\bm{x} \in \mathcal{X}}
		\frac{|A_k(\bm{x})|^2}{B_k(\bm{x})}.
		\label{eq:ratio_problem}
	\end{equation}
	\vspace{-8pt}
	\noindent is equivalent to
	\begin{equation}
			\max\limits_{\bm x \in \mathcal{X},\mu_k}\quad  2\mathcal{R}\left(\mu_k^* A_k\left(\bm x\right)\right) - |\mu_k|^2 B_k\left(\bm x\right),
			\vspace{-6pt}
	\end{equation}
	
	\noindent where $\bm \mu = [\mu_1,\mu_2,\dots,\mu_K]$.
	
	Proof: See {\rm~\cite{shen2018fractional}}. 
\end{theorem}

The SINR constraints in $\mathcal{C}_k$ introduce non-convexity with respect to the precoding vectors, preventing a direct convex solution of problem \eqref{Precoding Vector}.
Therefore, we reformulate this constraint by applying Theorem \ref{theorem1} as shown in the following corollary. 
\vspace{-6pt}
\begin{corollary}
	\label{corollary1}
	The constraint $\mathcal{C}_k$ can be reformulated as
	\vspace{-6pt}
\begin{equation*} 
	\label{FP} 
	\begin{aligned} 
		\xi \!\leq\! \max\limits_{\mu_{c,k}}
		2\mathcal{R}\{\mu_{c,k}^{*} \bm h_k^\mathrm{H} \bm w_c\}\!
		\!-\! |\mu_{c,k}|^2\!
		\left(\!
		\sum\nolimits_{i=1}^{K}{\!|\bm h_k^\mathrm{H} \bm w_i|^2}
		\!+\! \sigma_k^2\!
		\right)\!,\\
		\xi_k \!\leq\! \max\limits_{\mu_{k}}
		2\mathcal{R}\{\mu_{k}^{*} \bm h_k^\mathrm{H} \bm w_k\}\!
		\!-\! |\mu_{k}|^2\!
		\left(\!
		\sum\nolimits_{i=1,\,i \neq k}^{K}{\!|\bm h_k^\mathrm{H} \bm w_i|^2}
		\!+\!\sigma_k^2\!
		\right)\!.
	\end{aligned} 
\end{equation*}
	\vspace{-8pt}
\end{corollary}
\vspace{-8pt}

Then, we can acquire the optimal closed-form solutions of 
$\mu_{c,k}$ and $\mu_k$ as
\vspace{-8pt}
\begin{equation}
	\label{mu}
	\mu_{c,k}
	\!=\!
	\frac{\bm h_k^\mathrm{H}\bm w_c}
	{\sum_{i=1}^{K}\!|\bm h_k^\mathrm{H}\bm w_i|^2\!\!+\!\sigma_k^2},
	\mu_k
    \!=\!
	\frac{\bm h_k^\mathrm{H}\bm w_k}
	{\sum_{i=1,\,i\neq k}^{K}\!|\bm h_k^\mathrm{H}\bm w_i|^2 \!\!+\!\sigma_k^2}.
\end{equation}
\vspace{-12pt}

After the above transformation, the problem \eqref{Precoding Vector} is converted into a convex optimization problem that can be solved with a general optimization software toolbox, such as CVX.

\vspace{-10pt}
\subsection{Optimize Phase-Shift Matrix}
\vspace{-2pt}

With $\bm w$ fixed, we optimize $\mathbf{\Theta}$ by setting $\mathbf{\Theta}=\mathrm{diag}(\bm\theta^\mathrm{H})$. To decouple the coupled constraints, we introduce an auxiliary variable $\hat{\bm\theta}$ with the consistency constraint $\hat{\bm\theta}=\bm\theta$. Accordingly, problem (5b) can be equivalently reformulated as
	\begin{equation}
		\label{eq:split}
		\begin{aligned}
			\max_{\bm{\theta},\hat{\bm{\theta}},\xi,\xi_k}\    \mathcal{G} \quad		
			\text{s.t.}\  \mathcal{C}_{th},\ \mathcal{C}_k,\ \mathcal{C}_{\theta},\ \mathcal{C}_{\beta},
			\hat{\bm{\theta}}=\bm{\theta}.
		\end{aligned}
	\end{equation}
\vspace{-15pt}
\begin{corollary}
	\label{corollary2}
	Problem~\eqref{eq:split} can be reformulated using the augmented Lagrangian method as follows:
	\vspace{-5pt}
	\begin{equation}
		\label{Phase-Shift Matrix}
			\max\limits_{\bm\theta,\hat{\bm\theta},\xi,\xi_k} \ \mathcal{G}-\mu{\|{\hat{\bm \theta}-\bm \theta+\bm \nu}\|_2^2}, 
			\quad \text{s.t.} \ 
			\mathcal{C}_{th}, \mathcal{C}_k, \mathcal{C}_{\theta},\mathcal{C}_{\beta},
			\vspace{-9pt}
	\end{equation}

	\noindent where $\bm\theta \!=\! \left[\beta_1\!\left(\theta_1\right)e^{j \theta_1},\beta_2\!\left(\theta_2\right)e^{j \theta_2},\dots,\beta_N\!\left(\theta_N\right)e^{j \theta_N}\right]$, $\mu \!>\! 0$ is a given parameter, and $\bm \nu \in \mathbb{C}^{N \times 1}$ is a scaled dual variable.
\end{corollary}

The solution to problem \eqref{Phase-Shift Matrix} can be solved through the following updates
\vspace{-6pt}
\begin{subequations}
	\begin{flalign}
		\label{ADMM_theta}
		\bm \theta^{l+1}  &= \arg\max\limits_{\bm \theta}\quad\mathcal{G}-\mu{\|{\hat{\bm\theta}^l-\bm\theta+\bm\nu^l}\|_2^2}, \\
		\text{s.t.} \ &\mathcal{C}_{th} \!:\! \xi \geq \gamma_{th}, \xi_k \geq \gamma_{th}, \notag
		\label{ADMM_hat_theta}
		\mathcal{C}_k \!:\! \gamma_{c,k} \geq \xi ,\gamma_{k} \geq \xi_k, \notag\\
		\bm{\hat{\bm  \theta}}^{l+1}&=\arg\min\limits_{\hat{\bm \theta}}\quad{\|{ \hat{\bm\theta}-\bm\theta^{l+1}+\bm\nu^l}\|_2^2}, \\	
		\text{s.t.} \  & \mathcal{C}_{\theta}\!:\! -\pi \!\leq\! \hat{\theta}_n \!\leq\! \pi, \notag
		\mathcal{C}_{\beta}\!:\!		\beta_n(\hat{\theta}_n) \!=\! \eta \!\left( {\rm sin}(\hat{\theta_n} \!-\! \delta) \!+\! 1 \right)^{\alpha} \!\!+\! \beta_{\rm min},\notag\\
		\label{ADMM_nu}
		\bm\nu^{l+1}&=\bm\nu^l+\bm\hat{\bm \theta}^{l+1}-\bm\theta^{l+1}. 
	\end{flalign}
\end{subequations}
\vspace{-18pt}

The SINR conditions involved in \eqref{ADMM_theta} can be handled using the quadratic transformation proposed in Corollary 1.
Then, the constraint $\mathcal{C}_k$ can be equivalently reformulated by the following Corollary. 
\vspace{-6pt}
\begin{corollary}
	\label{corollary3}
	Define
	\vspace{-5pt}
	\begin{align}
		&\bm\Lambda_{c,k} = \mathrm{diag}\left(\bm f_k^\mathrm{H}\right)\mathbf{G} \bm w_c,\bm{\Lambda}_k = \mathrm{diag}\left(\bm f_k^\mathrm{H}\right)\mathbf{G} \bm w_k,\notag\\
		&\mathbf{P}_{c,k} = \sum_{i=1}^{K} \left( \mathrm{diag}\left( \bm f_k^\mathrm{H}  \right)\mathbf{G} \bm w_i \bm w_i^\mathrm{H} \mathbf{G}^\mathrm{H} \mathrm{diag}\left( \bm f_k \right) \right), \notag\\
		&\mathbf{P}_{k} = \sum_{i=1,i \neq k}^{K} \left( \mathrm{diag}\left( \bm f_k^\mathrm{H} \right)\mathbf{G} \bm w_i \bm w_i^\mathrm{H} \mathbf{G}^\mathrm{H}  \mathrm{diag}\left( \bm f_k \right) \right),\notag\\ 
		&h\! \left(\mu_{c, k},\bm \theta\right) \!=\! 2\mathcal{R}\!\left\{\mu_{c,k}^*\bm\Lambda_{c, k}^{H}\bm\theta\right\}\!-\!{|\mu_{c,k}|^2}\!\left(\bm\theta^\mathrm{H}\mathbf{P}_{c,k}\bm\theta \! + \! \sigma_k^2\right)\!, \notag\\
		&h\! \left(\mu_{k},\bm \theta\right) \!=\! 2\mathcal{R}\!\left\{\mu_{k}^*\bm\Lambda_{k}^{H}\bm\theta\right\} \! - \! {|\mu_{k}|^2}\!\left(\bm\theta^\mathrm{H}\mathbf{P}_{k}\bm\theta\!+\!\sigma_k^2\right)\!,\notag	
	\end{align}
	\vspace{-18pt}
	
	\noindent where $\mu_{c, k},\mu_{k}$ are calculated based on \eqref{mu}. 
	
	Constraint $\mathcal{C}_k$ is equivalent to the constraint
	\vspace{-8pt}
	\begin{equation}
		\begin{split}
			h\left(\mu_{c, k},\bm \theta\right) \geq \xi, \ 
			h\left(\mu_{k},\bm \theta\right) \geq \xi_{k}. 
		\end{split}
	\end{equation}
	\vspace{-16pt}

\end{corollary}
\vspace{-8pt}
Then, subproblem \eqref{ADMM_theta} is transformed into a convex problem, which can be solved effectively.

Furthermore, for problem \eqref{ADMM_hat_theta}, we propose a concise and effective solution method. Specifically, we first add $\mathcal{C}_\beta$ to the objective function to ensure that the amplitude-phase coupling constraint of practical RIS is satisfied. Then, problem \eqref{ADMM_hat_theta} can be decomposed into the following independent problem.
\vspace{-6pt}
\begin{equation}\label{line_search_method}
		\min_{\bm{\zeta}}  \
		 \left\|\bm{\zeta}
		\!-\!\left(\bm{\theta}\!-\!\bm{\nu}\right)\right\|_2^2  \ \ 
		\text{s.t.}\ 
		 \mathcal{C}_{\zeta}: 
		\zeta_n \!=\!\beta_n(\theta_n)e^{j\theta_n}, \mathcal{C}_{\theta},\ \mathcal{C}_{\beta}.
\end{equation}
\vspace{-15pt}


Since the objective and constraints in
\eqref{line_search_method} are separable across the RIS elements,
problem~\eqref{line_search_method} can be decomposed into $N$
independent element-wise problem, as follows
\vspace{-6pt}
\begin{equation*}
	\begin{aligned}
		\min_{-\pi\leq\theta_n\leq\pi}\quad
		& \beta_n^2(\theta_n)+|\vartheta_n|^2
		-2\beta_n(\theta_n)
		\left[
		{\rm Re}\{\vartheta_n\}\cos(\theta_n)
		\right.\\
		& \left.
		+{\rm Im}\{\vartheta_n\}\sin(\theta_n)
		\right].
	\end{aligned}
	\vspace{-6pt}
\end{equation*}

\noindent where $\vartheta_n=\theta_n^{\ell+1}-\nu_n^{\ell}$. 

Then, the optimal solution can be efficiently obtained by one-dimensional line search over feasible phase-shift range.

\begin{figure}[!t]
	\centering
	\subfloat[Convergence]{
		\includegraphics[width=0.47\linewidth]{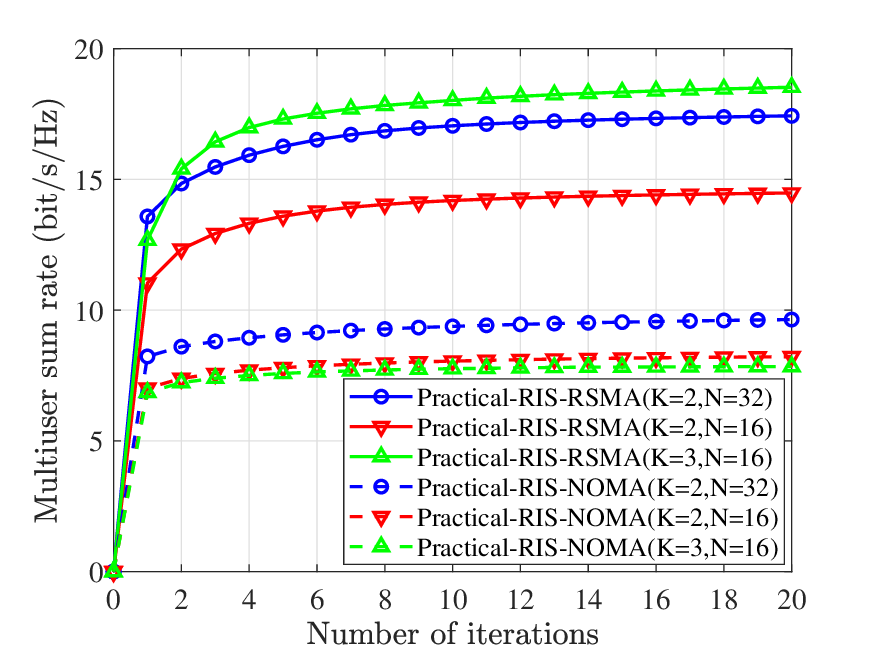}
		\label{fig2a}
	}
	\hspace{-3mm}
	\subfloat[RSMA vs. NOMA and OMA]{
		\includegraphics[width=0.47\linewidth]{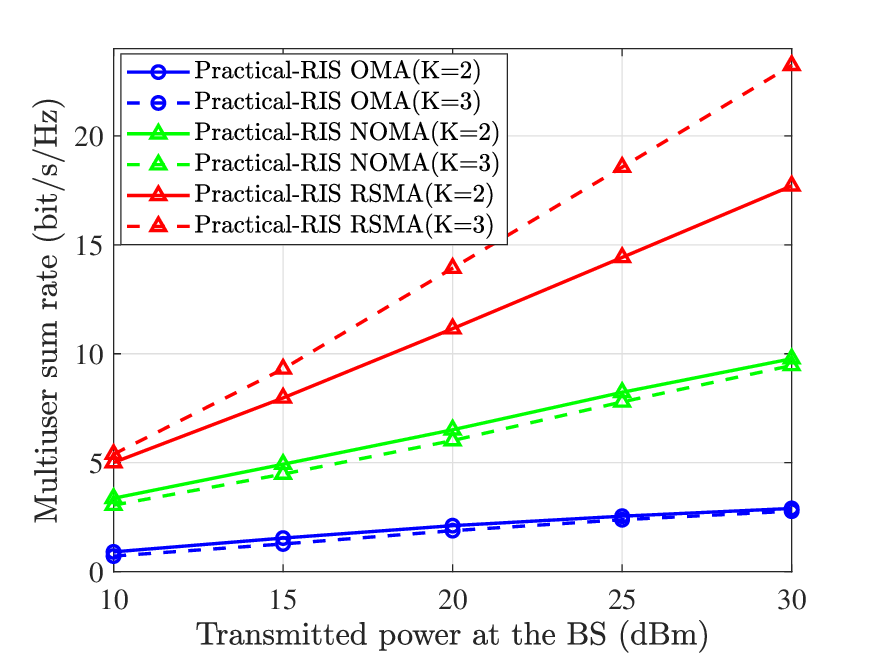}
		\label{fig2b}
	}
	\caption{Convergence and performance comparison of practical RIS-enhanced multiple-access systems.}
	\label{fig2}
	\vspace{-15pt}
\end{figure}

In addition, our proposed algorithm is fully compatible with 1-bit discrete phase-shift RIS reflection optimization. 
By simply adjusting the search range to $\left\{-\pi,0\right\}$, the optimization process can be effectively performed using the proposed algorithm. 
Moreover, in order to quantify the performance loss of projecting a continuous phase shift to a 1-bit discrete phase shift, we propose the following projection method. 
\vspace{-8pt}
\begin{equation}
	\begin{split}
		\Pi_{\mathcal{F}}\left(\theta_{n}\right) = \left\{
		\begin{array}{ll}
			(m-2)\pi, & \text{if } \theta_{n} < \frac{(2 m-3) \pi}{2}, \\
			(m-1)\pi, & \text{if } \theta_{n} \geq \frac{(2 m-3) \pi}{2}.
		\end{array}
		\right.
	\end{split}
\end{equation}
\vspace{-8pt}

\noindent where $m=\left\lfloor\frac{\theta_{n}+2\pi}{\pi}\right\rfloor$, and $\Pi_{\mathcal{F}}\left(\theta_{n}\right)$ is the projection of $\theta_n$.

\vspace{-6pt}
\section{simulation results}
\vspace{-3pt}
\label{s4}

In this section, we demonstrate the effectiveness of our algorithm. 
We set the BS at $(0,0,0)$m and the RIS at $(40,40,0)$m.
The users are distributed uniformly within a circle of radius 8 meters, centered at coordinates $(50, 0, 0)$m.
Rician fading is adopted to characterize the small-scale channel variations \cite{9520776}.
For both the BS-to-RIS and RIS-to-user links, the path loss is characterized as ${\rm PL}=37.3+22.0\log(d)$, with $d$ being the link distance.
Additionally, we set  $M \!=\! 8$, $N \!=\! 32$, $K \!=\! 2$, $\sigma_k^2\!=\!  -110$dBm, $\gamma_{th}\!=\! 1, \beta_{\min} \!=\! 0.2,\delta \!=\! 0.43\pi$, and $\alpha \!=\! 1.6$.

The convergence of the proposed algorithm is shown in Fig.~\ref{fig2}(a), where all simulation results are averaged over 200 independent experiments. As the number of iterations increases, the multiuser sum rate gradually improves and eventually converges.
Fig.~\ref{fig2}(b) compares the sum-rate performance of different multiple-access schemes versus the BS transmit power with practical RIS. As the transmit power increases, all schemes achieve higher sum rates, while the proposed RIS-enhanced RSMA scheme consistently outperforms the corresponding NOMA and OMA schemes. This advantage stems from RSMA's flexible interference management, which partially decodes interference through the common stream while treating the remaining interference as noise.

\begin{figure*}[!t]
	\centering
	\subfloat[{$P$}]{
		\includegraphics[width=0.26\textwidth]{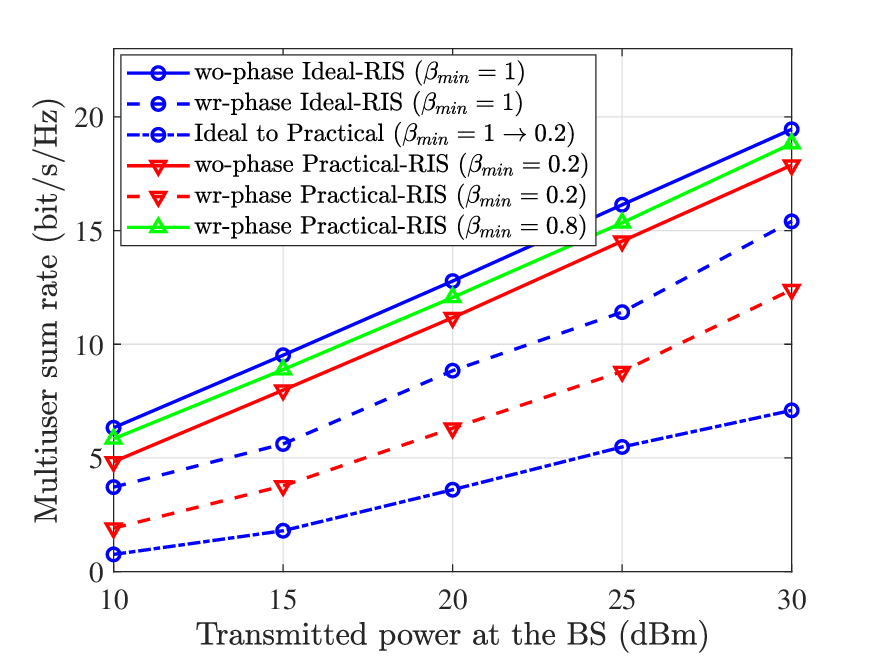}
		\label{fig3a}
	}
	\hspace{-0mm}
	\subfloat[{$M$}]{
		\includegraphics[width=0.26\textwidth]{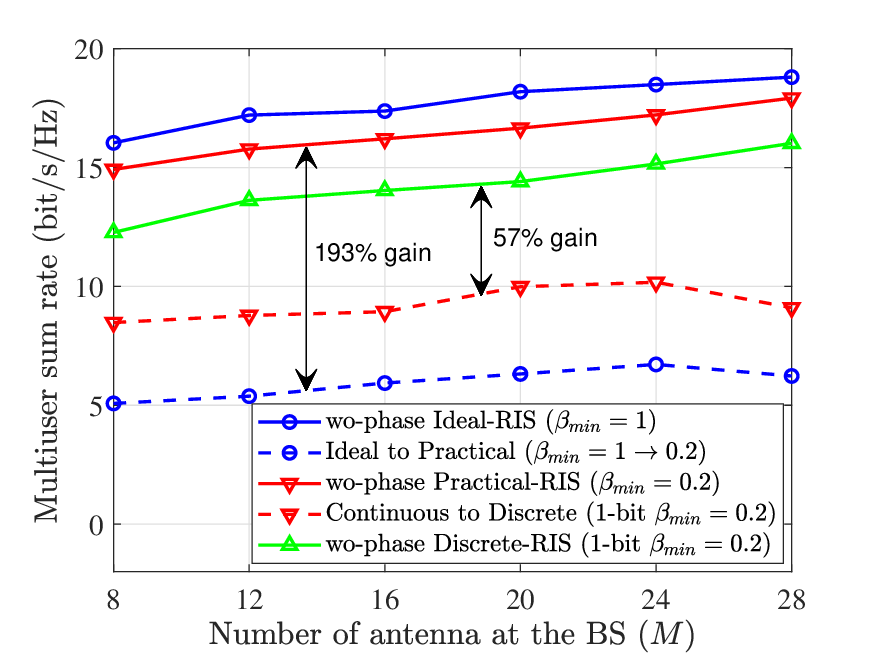}
		\label{fig3b}
	}
	\hspace{-0mm}
	\subfloat[{$N$}]{
		\includegraphics[width=0.26\textwidth]{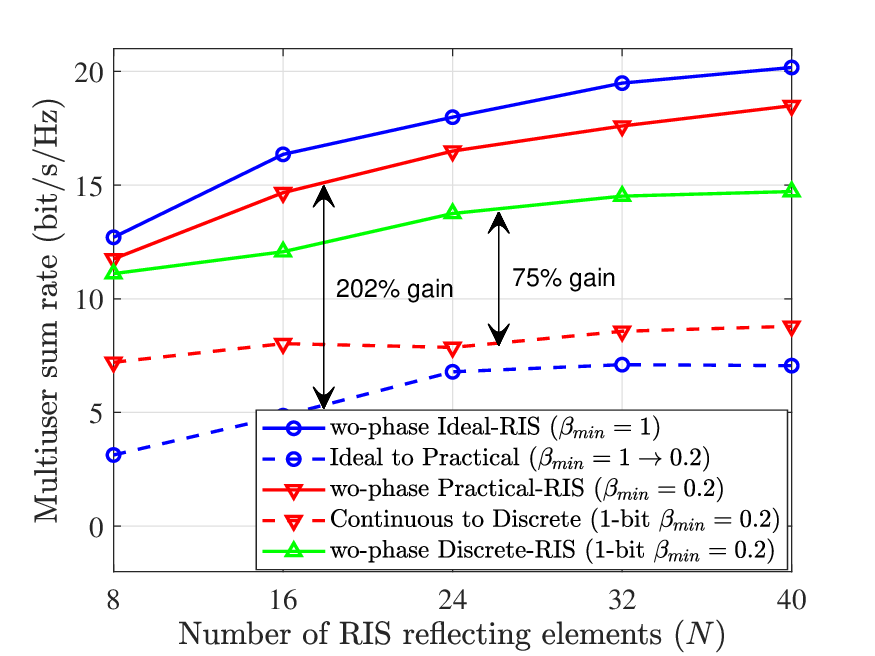}
		\label{fig3c}
	}
	\caption{The sum rate of practical RIS-enhanced systems under different parameters.}
	\label{fig3}
	\vspace{-18pt}
\end{figure*}


As shown in Fig.~\ref{fig3}, the ideal RIS (\(\beta_{\min}=1\)) achieves a higher sum rate than the practical RIS, indicating that reflection characteristics closer to those of the ideal RIS model can improve system performance. However, directly applying the phase shifts optimized under the ideal RIS model to a practical RIS system ("Ideal to Practical") results in significant performance degradation. In contrast, the proposed algorithm is applicable to both ideal and practical RIS models, demonstrating its good adaptability to different RIS hardware characteristics.
Moreover, the optimized phase-shift scheme (``wo-phase'') consistently outperforms the random scheme (``wr-phase''), validating the effectiveness of our algorithm. However, directly quantizing the phase shifts optimized under the continuous-phase-shift setting into 1-bit discrete values also causes noticeable performance degradation. The proposed algorithm, however, can directly optimize the discrete phase shifts, thereby effectively alleviating the performance loss caused by quantization.
As shown in Fig.~\ref{fig3}(b) and (c), the proposed scheme achieves sum-rate gains of approximately 200\% over the “Ideal to Practical” benchmark and improves the sum-rate by more than half compared with the continuous-phase-shift optimization followed by 1-bit quantization scheme.

\begin{figure}[!t]
	\centering
	\subfloat[{Ideal RIS}]{
		\includegraphics[width=0.47\linewidth]{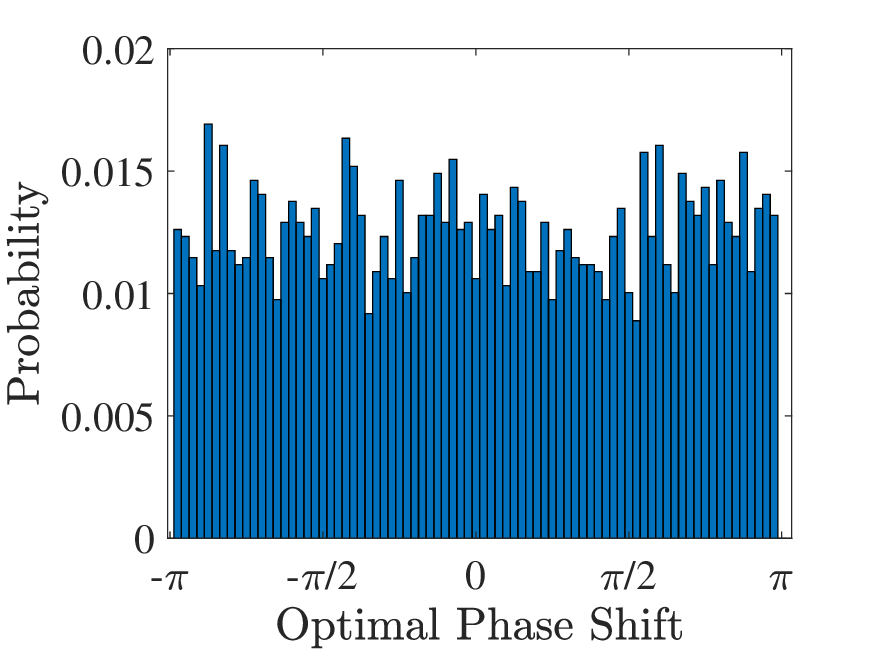}
		\label{fig4a}
	}
	\hspace{-3mm}
	\subfloat[{Practical RIS}]{
		\includegraphics[width=0.47\linewidth]{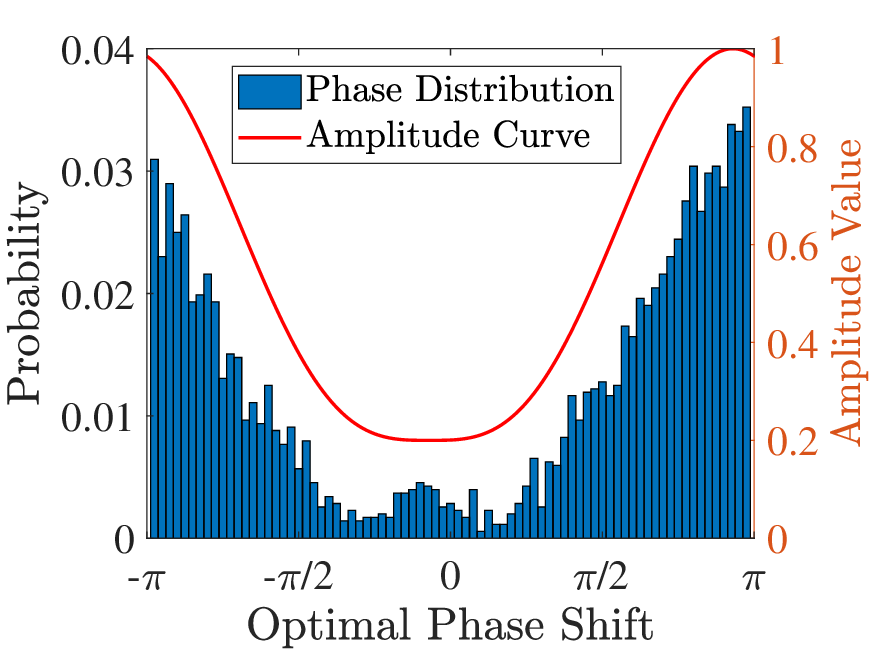}
		\label{fig4b}
	}
	\caption{Distribution of optimal phase-shift values for ideal and practical RIS.}
	\label{fig4}
	\vspace{-19pt}
\end{figure}

In addition, we investigate the optimal phase-shift distributions for the ideal RIS ($\beta_{\min}\!=\!1$) and the practical RIS ($\beta_{\min}\!=\!0.2$). As shown in Fig.~\ref{fig4}(a), the optimal phase shifts of the ideal RIS are approximately uniformly distributed over $\left[-\pi,\pi\right]$. In contrast, Fig.~\ref{fig4}(b) shows that the optimal phase shifts of the practical RIS are more concentrated around $\pi$ and $-\pi$, where the reflection amplitude approaches its maximum, while it is minimized at zero phase shift.

\vspace{-12pt}
\section{conclusion}

\vspace{-3pt}
\label{s5}
In this paper, we investigated practical RIS-enhanced RSMA systems, where the precoding vectors and phase shifts were optimized to maximize the multiuser sum rate. 
Specifically, we proposed an effective AO algorithm that was compatible with all the coupling relationships between the amplitude response and phase shift. 
In addition, the proposed algorithm had broad generality and could be directly applied to the multiuser sum-rate maximization problem for practical RIS-enhanced OMA or NOMA systems. 
Simulation results showed that the proposed scheme outperformed benchmark schemes in terms of system performance.

	\vspace{-5pt}
\bibliographystyle{IEEEtran}
\bibliography{myref}

\end{document}